\documentclass{article}

\usepackage[colorlinks=true,
            urlcolor=blue,
            linkcolor=docgreen,
            citecolor=docgreen,
            bookmarks=false,
            pdftitle={Comments on Lagrangian Transport by Breaking Surface Waves},
            pdfauthor={Douglas G. Dommermuth},
            pdfsubject={Energy Cascade},
            pdfkeywords={Ocean waves, Breaking waves,  Nonlinear Waves, wave growth, HOS, BWA},
	   bookmarksopen=false,
            pdfpagemode=UseNone]{hyperref}

\usepackage{natbib}
\usepackage{url}
\usepackage{hyperref}
\usepackage{ulem}
\usepackage{graphicx}
\usepackage{fixltx2e}
\usepackage{amssymb,amsmath}
\usepackage{dblfloatfix}
\usepackage[compact]{titlesec}
\usepackage{color}
\definecolor{docgreen}{rgb}{0,.5,0}
\usepackage[top=1in,bottom=1in,left=1in,right=1in,paperheight=11in,paperwidth=8.5in]{geometry}
\usepackage{pdflscape}
\usepackage{textcomp}
\usepackage{setspace}

\titlespacing{\section}{0pt}{*0}{*0}
\titlespacing{\subsection}{0pt}{*0}{*0}
\titlespacing{\subsubsection}{0pt}{*0}{*0}

\pdfoutput = 1

\newcommand\ie{i.e., }

\newcommand{\comment}[1]{}

\def\be{\begin{eqnarray}}
\def\ee{\end{eqnarray}}
\def\benl{\begin{eqnarray*}}
\def\eenl{\end{eqnarray*}}

\newcommand{\nwc}{\newcommand}
\nwc{\bm}{\boldmath}
\nwc{\m}{\mbox}
\nwc{\ubm}{\unboldmath}
\nwc{\bmU}{\m{\bm$U$\ubm}}
\nwc{\bmX}{\m{\bm$X$\ubm}}
\nwc{\bmu}{\m{\bm$u$\ubm}}
\nwc{\bmx}{\m{\bm$x$\ubm}}
\nwc{\bmz}{\m{\bm$z$\ubm}}
\nwc{\bmv}{\m{\bm$v$\ubm}}
\nwc{\bmw}{\m{\bm$w$\ubm}}
\nwc{\bmW}{\m{\bm$W$\ubm}}
\nwc{\bmn}{\m{\bm$n$\ubm}}
\nwc{\bmG}{\m{\bm$G$\ubm}}
\nwc{\bmF}{\m{\bm$F$\ubm}}
\nwc{\bmI}{\m{\bm$I$\ubm}}
\nwc{\bmN}{\m{\bm$N$\ubm}}
\nwc{\bmP}{\m{\bm$P$\ubm}}
\nwc{\bmcalP}{\m{\bm $\cal P$\ubm}}
\nwc{\bmV}{\m{\bm$V$\ubm}}
\nwc{\bmS}{\m{\bm$S$\ubm}}

\begin{document}
\addtocontents{toc}{\protect\setstretch{0.8}}

\title{Comments on the Wave Energy Cascade \\
{\normalsize Breaking Wave Analysis Technical Report No.\ 2 vers.\ 4}}

\author{Douglas G. Dommermuth \\
\href{mailto:Breaking\_Waves@cox.net}{Breaking\_Waves@cox.net} \\
%\href{http://doug.dommermuth.com}{http://doug.dommermuth.com} \\
Breaking Wave Analysis consulting \\
San Diego, CA  92103, USA}

\maketitle

\begin{center}
Abstract \\
\begin{quote}
The gravity-wave energy cascade is investigated using the High-Order Spectral (HOS) method. 
\end{quote}
\end{center}

\vspace{0.25in}

A JONSWAP spectrum is used to initialize the HOS simulations simulations of a seaway\citep{dommermuth1987}.   The wavelength at the peak of the spectrum ($L_o$) is used to normalize length scales.   The velocity scale is $U_o=\sqrt{g L_o}$.   Based on these choices for $L_o$ and $U_o$, the Froude number equals one ($F_r=1$). $k_o=2\pi$ is the wavenumber at the peak of the spectrum.  Details of the one-dimensional JONSWAP spectrum are provided in \cite{dommermuth2014}.  For the JONSWAP spectrum, the peak enhancement factor is $\gamma=3.3$, and the angular spreading is  $s=50$ with $\theta_o=0$ (see equations 51-54 of \cite{dommermuth2014}).    The HOS simulations use $N_x \times N_y = 1024 \times 256$ de-alaised Fourier modes with a fourth-order approximation.  The length (L), width (W), and depth (D) of the HOS simulations are respectively 20, 5, and 10.  The time step is $\Delta t=0.002$.   The linear portions of the HOS simulations are rolled up analytically (see equations 20-25 of \cite{dommermuth2014}).  The simulations are  adjusted with a period of adjustment $T_a=20$ (see equation 17 of \cite{dommermuth2014}).   The HOS simulations are smoothed every time step using bandpass filtering with $F_c=0.9$ (see equation 13 of \cite{dommermuth2014}), and each HOS simulation uses energy pumping to maintain the total energy (see 14-16 of \cite{dommermuth2014}) that would otherwise be lost through smoothing.   

\begin{figure*}
\begin{center}
\begin{tabular}{cc}
\includegraphics[angle=0,width=0.45\linewidth]{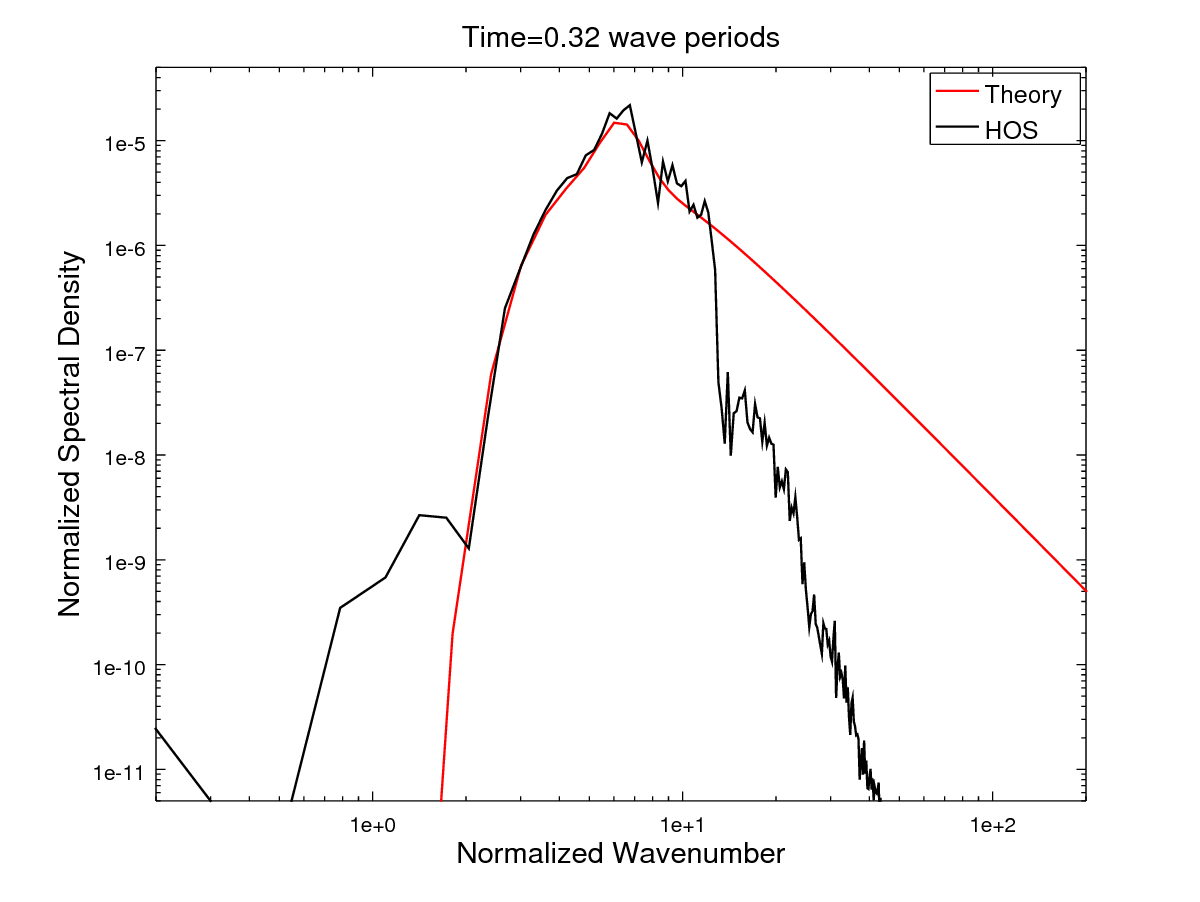} & 
\includegraphics[angle=0,width=0.45\linewidth]{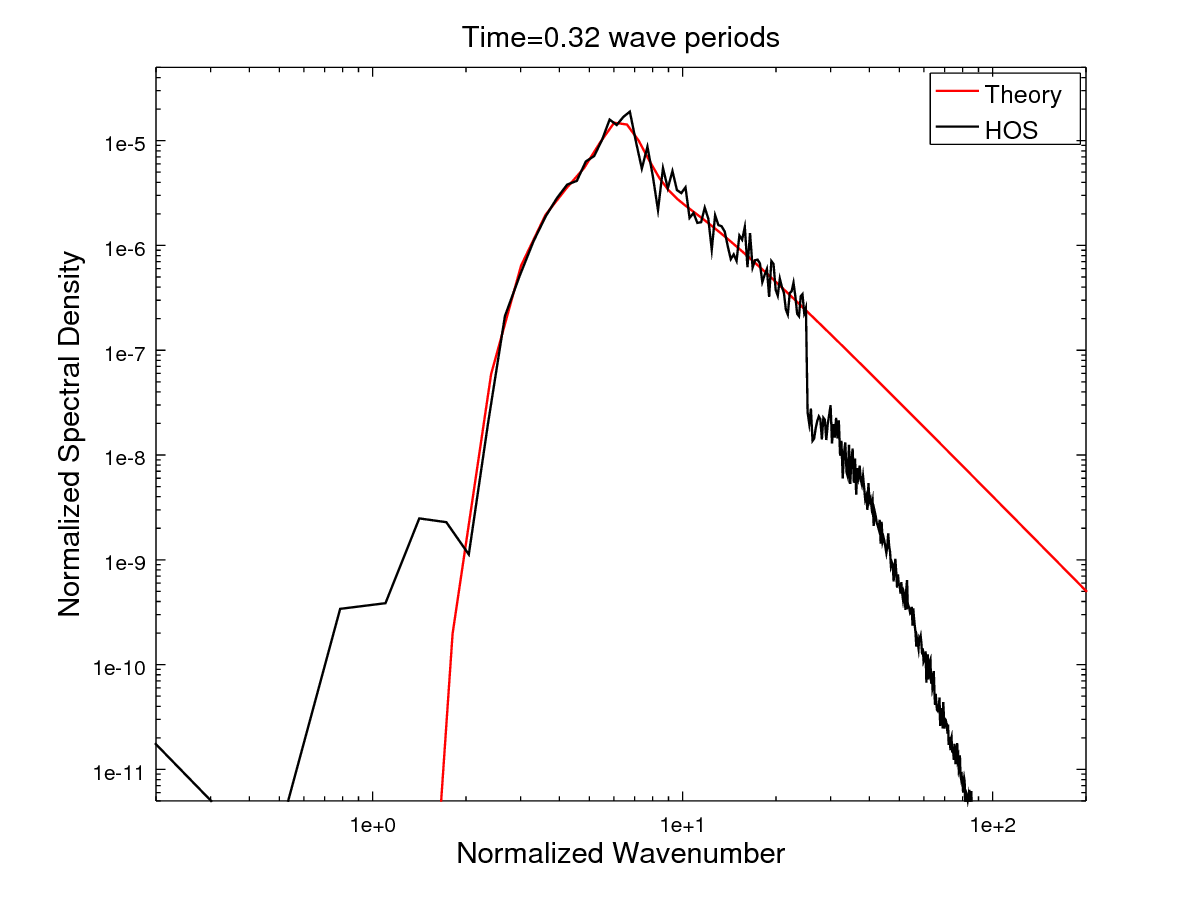}  \\
(a) & (b) \\
\includegraphics[angle=0,width=0.45\linewidth]{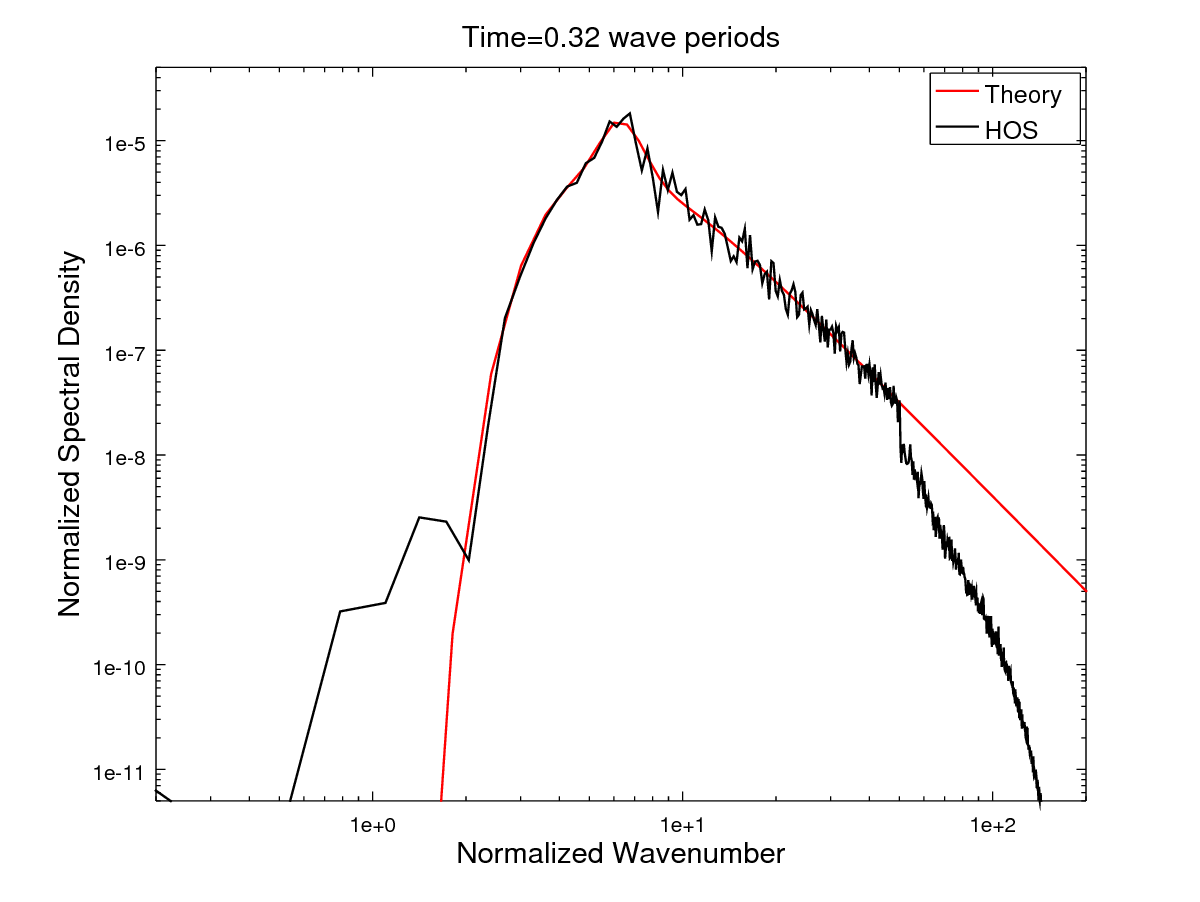} & 
\includegraphics[angle=0,width=0.45\linewidth]{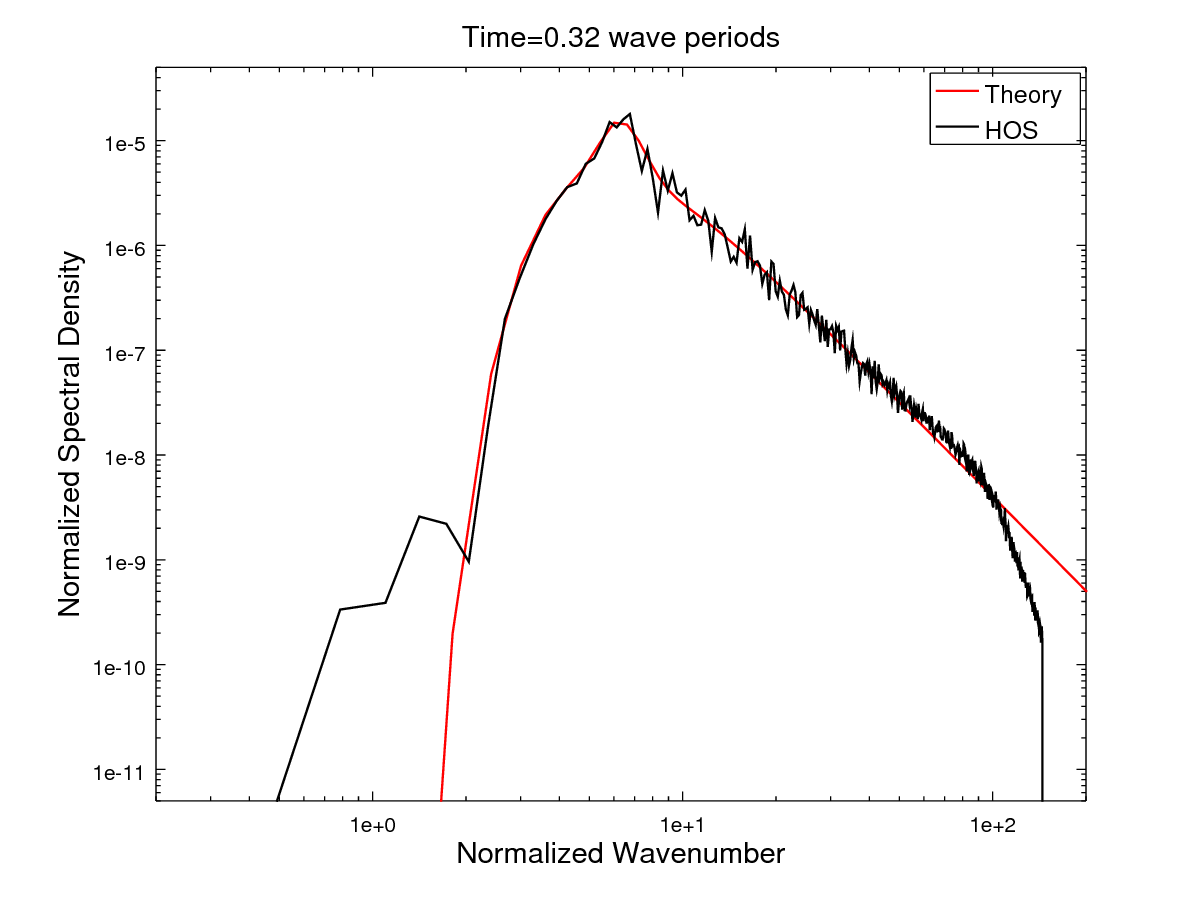}  \\
(c) & (d)
\end{tabular}
\end{center}
\caption{\label{fig:spec}  Wave Spectra.   Animations the evolution of the one-dimensional spectra as a function of time are available at
(a) \mbox{\cite{dommer2ko}} \href{https://youtu.be/KGZE3jsDSJ4}{$k_c=2k_o$};
(b) \mbox{\cite{dommer4ko}} \href{https://youtu.be/2Mjcq4H5neY}{$k_c=4k_o$};
(c) \mbox{\cite{dommer8ko}} \href{https://youtu.be/YzN4P4LSBrk}{$k_c=8k_o$}; and
(d) \mbox{\cite{dommer16ko}} \href{https://youtu.be/ULa0-CTsjd0}{$k_c=16k_o$}.}
\end{figure*}

A HOS simulation is run for 50,000 time steps to establish nonlinear interactions.  The resulting dataset is used to initialize four HOS simulations with varying amounts of bandpass filtering.   For these four HOS simulations, the Fourier content is cutoff for $k_c \geq 2 k_o, 4 k_o, 8 k_o$, and $16 k_o$.  Snapshots of animations of the one-dimensional energy spectra are provided in figure \ref{fig:spec}.  The snapshots are provided very early in the simulations.  The figure links to animations of the one-dimensional wave spectra as a function of time.  The green hyperlinks in the figure link to the references, and the blue hyperlinks connect to the YouTube animations.  The non-dimensional energy density is plotted on the y-axis, and the non-dimensional wavenumber is plotted on the x-axis.    The results of the HOS are compared to the theoretical JONSWAP spectrum.  The time in terms of the period of the wave at the peak of the spectrum ($T_o$) is shown at the top of the animations. 

There is no wind-forcing for the results that are shown in figure \ref{fig:spec}.   There is energy pumping to compensate for any energy that is lost through smoothing, but I think that its effects are minimal.   For all of the simulations, a $k^{-3}$ power-law fills in over time for $k > k_o$.   Similar results are reported in \mbox{\cite{dommermuth2010a}}, \mbox{\cite{dommermuth2013}}, and \mbox{\cite{dommermuth2014}}.   For $k>k_o$, locked-wave modes fill in immediately.  The locked-wave modes are slowly converted into free-wave modes over time.  The $k^{-3}$ power-law fills in for $k_c \geq 2 k_o, 4 k_o, 8 k_o$, and $16 k_o$ at the approximate rate of 1150, 350, 120, and 25 peak wave periods ($T_o$), respectively.   As the cut-off wavenumber is cut in half, the time that it takes the spectrum to saturate is almost quadrupled, which suggests an inverse quadratic relationship.  Although the wave numbers near $k_o$ are not resolved well because the length and width of the domain are too short, there is some evidence of down shifting toward lower wave numbers of the spectra for $k < k_o$.  These simple numerical experiments also lead to several interesting observations:

\begin{enumerate}
\item The mechanism for Fermi-Pasta-Ulam recurrence is broken for broad-banded wave spectra with an extensive $k^{-3}$ power-law region.    The upper and lower side bands do not show periodicity because band-passed wave spectra fill in with a $k^{-3}$ power-law.

\item As locked-wave modes are converted into free-wave modes, the most unstable modes must occur at very specific phases up and down the wave spectrum.   This suggests that the phases of free-wave modes are interrelated throughout the wave spectrum.   It is no wonder that \cite{dommermuth2013} shows that extreme events are very sensitive to phase perturbations.

\item  The wave energy moves toward higher wavenumbers toward where parasitic capillary waves come into play \citep{dommermuth1994a,mui1994}.
Interestingly, the phases of the parasitic capillaries are determined by the underlying gravity wave, so in some sense the phases of the parasitic waves can be tied back all the way to the phases of the waves in some small neighborhood at the peak of the spectrum.   For $k>k_o$, what is not obvious about the current numerical experiments is that there is also back scatter of energy from high to low wavenumbers, \ie downshifting.   Energy is also moving up the spectrum from the parasitics to the peak of the spectrum at $k_o$.   The current numerical experiments emphasize the forward scattering of wave energy.

\item Wind input is not required to move wave energy off the peak of the spectrum.   In fact, the wave energy can only move off the peak at a finite rate due to nonlinear wave interactions.   For growing seas, the peaks in the wave spectra that are observed are due to this finite rate of energy exchange.   The energy exchange associated with the action of wind must enter the wave field in a small neighborhood of the peak of the wave spectrum.  Off the peak, the waves act as turbulent roughening for the wind and wind-wave drift in the water.    As noted by \cite{dommermuth2014}, the normal stress due to the drift velocity in the water needs to be considered in addition to the normal stress due to the wind. I think that the shorter waves off the peak of the wave spectrum drive the turbulence in the air and the water rather than the other way around.
\end{enumerate}

This research could be extended by performing numerical simulations with longer and wider domains while keeping the same resolution at high wavenumbers.   Under such circumstances, the approximations in the HOS method begin to break down as shown by \cite{dommermuth2010a}.     \cite{dommermuth2010a} shows that Free-Surface Mapping (FSM) permits simulations of broad-banded wave spectra that are difficult to investigate using HOS.   Other candidates are the volume-of-fluid (VOF) schemes that are used in \cite{dommermuth2010b}, \cite{dommermuth2013}, and \cite{dommermuth2014}, but VOF is computationally much more intense than FSM.   \cite{dommermuth1994a} compares the advantages and disadvantages of two different fourth-order FSM solvers.  \cite{dommermuth1994a} uses the formation of parasitic capillary waves to illustrate one application of FSM that is impossible using HOS.  FSM could also be used to look at swell interacting with wind waves, which would be very difficult using HOS.

Another key advantage of FSM compared to HOS is that it provides the solution to the flow field in the water and by extension, the air.   \cite{dommermuth1993} shows how to couple the vortical and wavy portions of the flow using an Helmholtz decomposition.    \cite{dommermuth1993} uses FSM with hybrid finite-difference and spectral methods to investigate the interaction of a vortex pair with a free surface.   The interaction with the wavy portion of the flow occurs through a particular component of the pressure that is associated with the vortical portion of the flow.   \cite{dommermuth1993} shows convergence using fourth and sixth-order finite-difference schemes based on FSM methods.   \cite{dommermuth1994b} shows that the free-surface elevation is hydrostatically balanced with the vortical component of pressure at low turbulent Froude numbers. My point here is that researchers need to consider the vortical components of the pressure in the air and the water in their studies of wind-wave growth.   These studies can be accomplished by coupling a Helmholtz decomposition of the flow in the air with a Helmholtz decomposition of the flow in the water. I think that it is possible to investigate broad-banded wave spectra and wind-wave growth simultaneously by performing this coupling.  My own research interests are going down this path wherein I consider the how the largest temporal and spatial scales in the wind in the air and wind-drift in the water affect the growth of waves at the peak of the wave spectrum.  For the shorter waves off the peak of the spectrum, I think that spatial and temporal scales of the waves relative to the scales of the turbulence in the air and water are too disparate for there to be a meaningful transfer of energy.  The current research shows that energy input at the peak of the spectrum is distributed through nonlinear wave interactions to the entire wave spectrum.    As constructed, the current numerical experiments tend to emphasize the forward scattering of wave energy from low to high wavenumbers.   There is also backward scattering of wave energy that is occurring up and down the spectrum. 

\section*{Acknowledgement}

This research is self-funded by Breaking Wave Analysis consulting.  (Breaking Wave Analysis is a S corporation in the state of California.) My current research is focused on modeling broad-banded wave spectra and wind forcing at the peak of the spectrum.   If you are interested in funding my research, contact me at  \href{mailto:Breaking\_Waves@cox.net}{Breaking\_Waves@cox.net}.  I am particularly interested in collaborating with wind-wave tanks.   The objective of such collaborations would be to assimilate experimental measurements into numerical simulations of wind-driven waves.   For the price of a latte, you can also support my research by sending money through PayPal to \href{mailto:Breaking\_Waves@cox.net}{Breaking\_Waves@cox.net}.

\bibliographystyle{dommerbib}
\bibliography{dommerbib}

\end{document}